\documentclass[10pt]{iopart}

\usepackage{amssymb}
\usepackage{graphicx}
\usepackage{xcolor}
\usepackage[colorlinks=true]{hyperref}

\begin{document}

\title[Assessing the Probability of Extremely Rare Low Wind Energy Production in Europe]{Assessing the Probability of Extremely Low Wind Energy Production in Europe at Sub-seasonal to Seasonal Time Scales}

\author{Bastien Cozian$^1$, Corentin Herbert$^2$ and Freddy Bouchet$^{2,3}$}

\address{1. ENS de Lyon, UCBL, CNRS, Laboratoire de physique, F-69342 Lyon, France}
\address{2. ENS de Lyon, CNRS, Laboratoire de physique, F-69342 Lyon, France.}
\address{3. LMD/IPSL, ENS, Université PSL, École Polytechnique, Institut Polytechnique de Paris, CNRS Paris, France}
\ead{bastien.cozian@ens-lyon.fr}
\vspace{10pt}

\begin{abstract}
The European energy system will undergo major transformations in the coming decades to implement mitigation measures and comply with the Paris Agreement.
In particular, the share of weather-dependent wind generation will increase significantly in the European energy mix.
The most extreme fluctuations of the production at all time scales need to be taken into account in the design of the power system.
In particular, extreme long-lasting low wind energy production events constitute a specific challenge, as most flexibility solutions do not apply at time scales beyond a few days.
However, the probability and amplitude of such events has to a large extent eluded quantitative study so far due to lack of sufficiently long data.
In this letter, using a 1000-year climate simulation, we study rare events of wind energy production that last from a few weeks to a few months over the January-February period, at the scale of a continent (Europe) and a country (France).
The results show that the fluctuations of the capacity factor over Europe exhibit nearly Gaussian statistics at all time scales.
A similar result holds over France for events longer than about two weeks and return times up to a few decades.
In that case, the return time curves follow a universal curve.
Furthermore, a simple Gaussian process with the same covariance structure as the data gives good estimates of the amplitude of the most extreme events.
This method allows to estimate return times for rare events from shorter but more accurate data sources.
We demonstrate this possibility with reanalysis data.
\end{abstract}

\vspace{2pc}
\noindent{\it Keywords}: rare events, wind energy production, energy meteorology \\
\maketitle
\ioptwocol
%

%%%%%%%%%%%%%%%%%%%%%%%%%%%%%%%%%%%%%%%%%%%%%%%%
% Introduction
%%%%%%%%%%%%%%%%%%%%%%%%%%%%%%%%%%%%%%%%%%%%%%%%
\section{Introduction}
% Variability of renewable energy on various timescales,
Over the past decades, the share of renewable energy in the energy mix of many European countries has been rapidly increasing.
This trend should only intensify in the coming years, as rapid deployment of renewable energy is an integral part of the global efforts to contain global warming within the bounds of the Paris agreement~\cite{Rogelj2015,vanVuuren2018}.
As a consequence, electricity production is becoming increasingly sensitive to weather conditions~\cite{Bloomfield2016,Bloomfield2018}.
One of the main challenges of the energy transition is to maintain the balance between electricity production and demand in spite of this increased variability.
This challenge is actually a multi-facet one, as the energy production fluctuations occur on a wide spectrum of timescales.
For instance, photovoltaic production might fluctuate on a timescale on the order of minutes due to local variations of cloud coverage and surface solar irradiance~\cite{Tomson2006, Lave2012, Watanabe2016}, and wind energy also exhibits variability in this time range~\cite{Holttinen2005, Apt2007, Anvari2016}.
But these energy sources, and in particular wind energy, also fluctuate on longer timescales, due to the prevalence of different meteorological conditions on timescales of days, weeks, months, or even years.
% The specific challenge of sub-seasonal to seasonal timescales
In this letter, one of our main goals is to estimate the probability, or \emph{return time}, and amplitude of events where wind energy production remains anomalously low for periods ranging from a few days to the whole winter.
In particular, we focus on longer events lasting several weeks, because they impose specific constraints on the design of the energy system.
Indeed, many standard flexibility solutions (e.g. demand-side management, battery storage or even pump storage power plants) become unavailable on timescales exceeding a few days, let alone a whole season.
Hence, even if they should be expected to be lower than for shorter events, quantifying precisely the needs for flexibility and backup on such long timescales is crucial to ensure the security and robustness of future energy systems, and to assess the corresponding economic cost.

% State of the art
Variability at the sub-seasonal to seasonal timescales has attracted growing interest over the past years.
For instance, low wind speed events lasting up to 20 days have been studied over Ireland~\cite{Leahy2013} and over the North Sea~\cite{Patlakas2017}, using 30-year records from weather stations and a 10-year high-resolution simulation from a regional model, respectively.
Similar studies, relying on reanalysis data, have also been conducted for low wind power events in Germany~\cite{Jung2018, Ohlendorf2020}, and Great Britain~\cite{Cannon2015}.
% Rare events
However, a fundamental challenge remains, as estimating the characteristics of such events in a statistically robust way requires much more data than is available from observations.
Indeed, extreme events play a key role due to their major socioeconomic impact, but require time series much longer than their return time.
Estimating the probability and characteristics of rare, long-lasting low anomalies of energy production thus requires new methodological progress.

% This paper
In this letter, we address this methodological question, and quantify the return time and amplitude of low extremes of wind energy production.
We show that return times for events lasting from a few days to the whole winter can be obtained from long simulations with climate models, here 1000 years of data, using a simple energy model to relate energy production to climate variables similar to the one used by van der Wiel et al.~\cite{vanderWiel2019a,vanderWiel2019b} to study the meteorological conditions associated with low production and high demand events over Europe.
The first main novelty of this work is the computation of return levels associated with return times up to several centuries, and the discussion of their dependence on different scenarios of installed capacity and on various time and space aggregation scales.
In particular, we show that for long aggregation times, the capacity factor fluctuations follow a universal curve, independent of the scenario.
We further discuss the impact of the duration of the event and the effect of spatial aggregation.
For instance, for 20-year events, relative fluctuations of capacity factor for short events (on the order of days) are close to 100\% at the scale of France, and 80\% at the scale of Europe, while they reduce to about 30\% for France and 20\% for Europe for two-month events.
Such fluctuations still represent a wind power shortfall on the order of 10GW for France and 60GW for Europe for the whole two-month period, which might prove challenging in terms of flexibility and backup.
The second main contribution of this work is to show that return times for low wind energy production can be estimated from a simple Gaussian process, whose parameters are estimated from available data.
We show that this method provides a good approximation for large spatial scales (here Europe) at any time scale, or at smaller spatial scale (France) for events longer than about two weeks and return times up to a few decades.
As an application, we estimate return times from reanalysis data for events which are much more rare than would be possible using only the raw data.

%%%%%%%%%%%%%%%%%%%%%%%%%%%%%%%%%%%%%%%%%%%%%%%%
% Section 1
%%%%%%%%%%%%%%%%%%%%%%%%%%%%%%%%%%%%%%%%%%%%%%%%

\section{Data and methods}

\subsection{Climate and wind energy model}

Following the same strategy as other studies based on reanalysis~\cite{MacLeod2018,Bloomfield2021c} or climate models~\cite{vanderWiel2019a}, we construct time series of wind energy production from climate variables, taken from a 1000-year statistically stationary simulation with 3-hourly output using the CESM model~\cite{Hurrell2013}.
Our setup uses atmosphere and land components only.
The sea surface temperature, sea ice cover and greenhouse gases concentration are prescribed at values representative of the 2000s.
The same dataset was used recently to study heat waves over Europe~\cite{Ragone2021}.
From the 10m-wind speed in the climate model output, we compute the wind speed at turbine height $\mathrm{WS}$, for each grid point $(x,y)$ and time $t$.
The wind speed is then converted to a local capacity factor (ratio of the generated power to its nominal value) through a power curve $\mathrm{CF}(x,y,t)=\mathrm{PC}[\mathrm{WS}(x,y,t)]$.
The power curve we use is a simple 3-parameter model based on the features of industrial wind turbine power curves~\cite{Mathew2006, Troccoli2018} (\ref{sec:windmodelappendix}).

As we are interested in persistent extreme events, we look at the effect of time aggregation by computing the running average
\begin{equation}
\mathrm{CF}_{T}(t) = \frac{1}{T}\int_t^{t+T} \mathrm{d}u\,\mathrm{CF}(u),
\label{eq:CF_T}
\end{equation}
where $T$ is the integration time.
We consider the time series $\mathrm{CF}(t)$ and $\mathrm{CF}_T(t)$ as realizations of a periodic stochastic process (with 1-year period), and we denote $\langle \cdot \rangle$ the average over realizations, estimated from the empirical average over the 1000 years of data.
We shall also study the statistics of the yearly lowest capacity factor event aggregated over some time period $T$ (\ref{sec:returntimesappendix}).

\subsection{Scenarios of installed capacity}\label{sec:scenarios}

The total capacity factor over any region $\mathcal{D}$ is computed as
\begin{equation}
\mathrm{CF}(t) = \int_{\mathcal{D}} \mathrm{d}x\mathrm{d}y \frac{\mathrm{IC}(x,y)}{\mathrm{IC}_{\mathcal{D}}}\mathrm{CF}(x,y,t),
\label{eq:CF}
\end{equation}
where $\mathrm{IC}(x,y)$ (in $\mathrm{W.m^{-2}}$) and $\mathrm{IC}_{\mathcal{D}}$ (in $\mathrm{W}$) are the local and total installed capacity, respectively.
We use the installed capacity of five scenarios developed by the e-Highway2050 project\footnote{Documentation available at \url{https://docs.entsoe.eu/baltic-conf/bites/www.e-highway2050.eu/results}.}~\cite{eHighway}.
This project provides distributions of onshore and offshore wind installed capacity based on current and future plans as of 2015.
More recent scenarios in line with the Paris Agreement and consistent with the European Green Deal have since been published~\cite{REF2020,TYNDP2022,REF2016,TYNDP2020}.
However, the total installed capacity and spatial distribution are more contrasted in the e-Highway2050 scenarios.
This allows us to explore the influence of a wider range of choices on extreme events, and to draw more robust conclusions likely to apply to other scenarios as well.

e-Highway2050 provides installed capacities at the scale of ``clusters'' (Fig.~\ref{fig:spatial_distribution_IC_eHighway}), which we aggregate over two domains to study the effect of space aggregation on extremes:  $\mathcal{D}=$ Europe, or $\mathcal{D}=$ France.
Because e-Highway2050 assumes that all offshore capacity is located in the North Sea, our Europe region is defined as the onshore capacity for 34 country plus the offshore capacity in the North sea, while our France region contains only onshore capacity.
Table~\ref{tab:eHighway} shows the total installed capacity for each scenario in each region.
The total installed capacity varies from 304 GW to 875 GW.
The share of offshore capacity is around 15\% in all scenarios, except for scenario X16 where it is significantly lower (4\%).

\begin{table}
\caption{\label{tab:eHighway}Total installed capacity of 34+1 countries in Europe (i.e. onshore capacity of 34 countries and offshore capacity in the North Sea), onshore capacity in France, and offshore capacity in the North Sea for the 5 e-Highway2050 scenarios. The share of offshore capacity, exclusively located in the North Sea, is shown in parenthesis.}
\begin{indented}
\lineup
\item[]\begin{tabular}{@{}llll}
\br
Scenario & Europe & France & North Sea \\
\mr
X5       & 813 GW & \084 GW &  104 GW (13\%) \\
X7       & 875 GW &  124 GW &  115 GW (13\%) \\
X10      & 512 GW & \058 GW & \076 GW (15\%) \\
X13      & 304 GW & \022 GW & \042 GW (14\%) \\
X16      & 388 GW & \054 GW & \015 GW (4\%) \\
\br
\end{tabular}
\end{indented}
\end{table}

%%%%%%%%%%%%%%%%%%%%%%%%%%%%%%%%%%%%%%%%%%%%%%%%
% Section 2
%%%%%%%%%%%%%%%%%%%%%%%%%%%%%%%%%%%%%%%%%%%%%%%%

\section{Capacity factor statistics in Europe and France}
\label{sec:cf_statistics}

\subsection{Annual mean capacity factor}
\label{sec:cf_annual}

The annual mean capacity factor $\langle \mathrm{CF}_{T=1y} \rangle$ in our model ranges from $25.5\%$ for the X16 scenario to $28.9\%$ for the X13 scenario.
% Comparison with van der Wiel
Van der Wiel et al.~\cite{vanderWiel2019a} found significantly smaller values: using $2000$ years of data representative of present-day climate, for two different climate models (EC-Earth and HadGEM2-ES), the average daily wind energy production they report (2.1 TWh/day and 1.3 TWh/day, respectively) correspond to annual mean capacity factors of $18.3\%$ and $11.3\%$, respectively.
This discrepancy can be explained by several factors: a smaller set of 15 European countries, and different parameters for the wind energy and climate models.
Using their wind energy model parameters, the same set of 15 countries and the same scenario (X5), we find an annual mean capacity factor of $19.3\%$ with our climate model.
The rest of the difference can be attributed to the climate data.
With the same installed capacity but using ERA interim data leads to an annual mean capacity factor of $22.6\%$~\cite{vanderWiel2019a}.
% Comparison with Renewables.Ninja dataset (from reanalyses)
Similarly, the Renewables.Ninja dataset of capacity factor timeseries calculated with MERRA-2 reanalysis leads to an annual mean capacity factor of $24.2\%$ with the actual 2017 installed capacity of 24 European countries~\cite{Staffell2016}.
This same dataset yields an annual mean capacity factor of $26.3\%$ with the X5 scenario over the same countries.

Hence, while the annual mean capacity factor given by such energy models depends on many parameters which make precise comparison difficult, our values are broadly compatible with the existing literature.
Here we will focus on fluctuations of the capacity factor rather than its absolute value.

\subsection{Seasonal cycle of capacity factor}
\label{sec:cf_seasonal}

\begin{figure*}[htbp]
\centering
\includegraphics[width=0.5\textwidth]{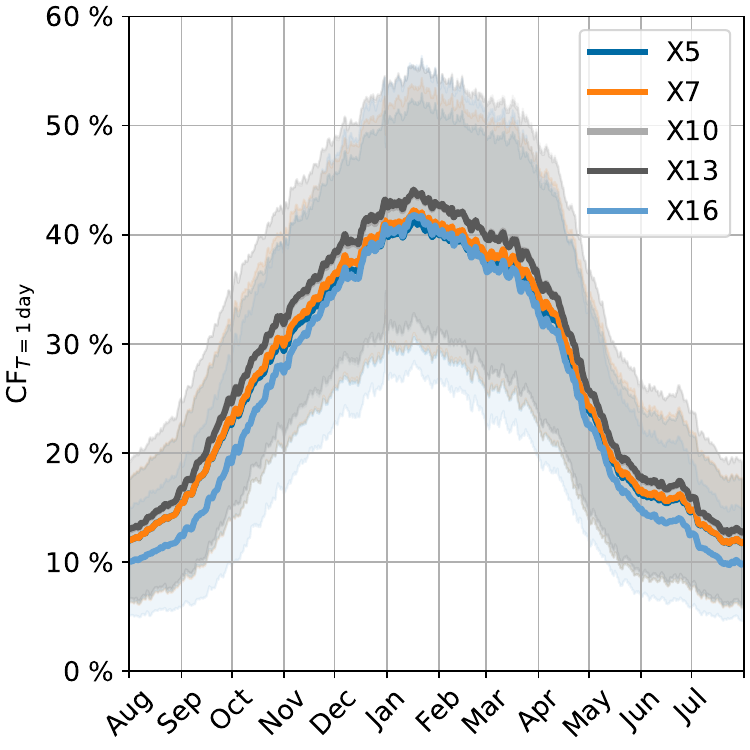}
\caption{Daily capacity factor $\mathrm{CF}_{T=1\,\mathrm{day}}$ in Europe: average (solid lines) and standard deviation (shaded area) over 1000 sample years, for 5 e-Highway2050 scenarios of installed capacity.
The curves for scenarios X10 and X13, and to a lesser extent X5 and X7, are indistinguishable from each other.
}
\label{fig:mean_std_annual_CF}
\end{figure*}
Wind power has a strong seasonal cycle.
Figure~\ref{fig:mean_std_annual_CF} shows the statistics of the daily capacity factor $\mathrm{CF}_{T=1\,\mathrm{day}}$ in Europe for the five scenarios.
We find that it is about $2.9$ times larger in winter ($40.2\%$ in DJF, averaged over the 5 scenarios) than in summer ($13.9\%$ in JJA).
This is consistent with~\cite{vanderWiel2019a} who find a winter-summer ratio of $2.5$ and $3.5$ for their two climate models, but larger than the value they report for ERA-Interim, which is 2.2.

We will focus here on the winter season, when extremely low wind generation events can be expected to have the largest impact.
Indeed, it is the time of the year where both the energy demand and fluctuations of wind energy capacity factor are largest (figure~\ref{fig:mean_std_annual_CF}).
Specifically, we will consider events occurring in January and February, because this period is short enough to consider the time series of capacity factor to be approximately stationary (the mean and variance of the daily capacity factor in figure~\ref{fig:mean_std_annual_CF} are approximately constant) and still long enough to provide meaningful statistics (e.g. time correlations).
Hence, all the winter-averages in the sequel correspond to JF averages.

As Fig.~\ref{fig:mean_std_annual_CF} shows, the seasonal capacity factor statistics vary from one scenario to the next.
In particular, a higher share of offshore capacity decreases the variability of wind production in our model, in agreement with previous studies~\cite{Drew2015} (see~\ref{sec:windmodelappendix} for more details).

\subsection{Effect of space and time aggregation on capacity factor statistics}

The smoothing effect on the variability of aggregating wind energy production from distant areas has been documented in a number of studies based on observational data~\cite[e.g.]{Archer2007,Kempton2010,Hasche2010,Fertig2012,Louie2014,Handschy2017}.
Here, we also observe this effect: while the PDF of winter capacity factor at the scale of France is strongly skewed, it becomes close to Gaussian at the scale of Europe, and the variance is reduced by a factor 2.

Similarly, fluctuations on different timescales, characterized by different solutions in terms of backup and flexibility, have different statistical properties.
At the elementary level, one can note that aggregating wind energy production over longer and longer timescales makes it more and more Gaussian and reduces its variance, similarly to aggregation in space.

Both these properties are illustrated in~\ref{sec:aggregation_statistics}.
Now, we turn to the main point of our paper, which is to discuss quantitatively the smoothing effect due to space and time aggregation on rare fluctuations of the capacity factor, rather than typical ones.

%%%%%%%%%%%%%%%%%%%%%%%%%%%%%%%%%%%%%%%%%%%%%%%%
% Section 3
%%%%%%%%%%%%%%%%%%%%%%%%%%%%%%%%%%%%%%%%%%%%%%%%

\section{Return time curves}\label{sec:return_times}

%%%%%%%%%%%%%%%%%%%%%%%%%%%%%%%%%%%%%%%%%%%%%%%%

\subsection{Return times of winter-average capacity factor}

In this section, we assess the probability of low winter wind energy production events by computing return time curves~\cite{Gumbel1941,Lestang2018} (see~\ref{sec:returntimesappendix} for details).

\begin{figure*}[htbp]
\centering
\includegraphics[width=0.3\textwidth]{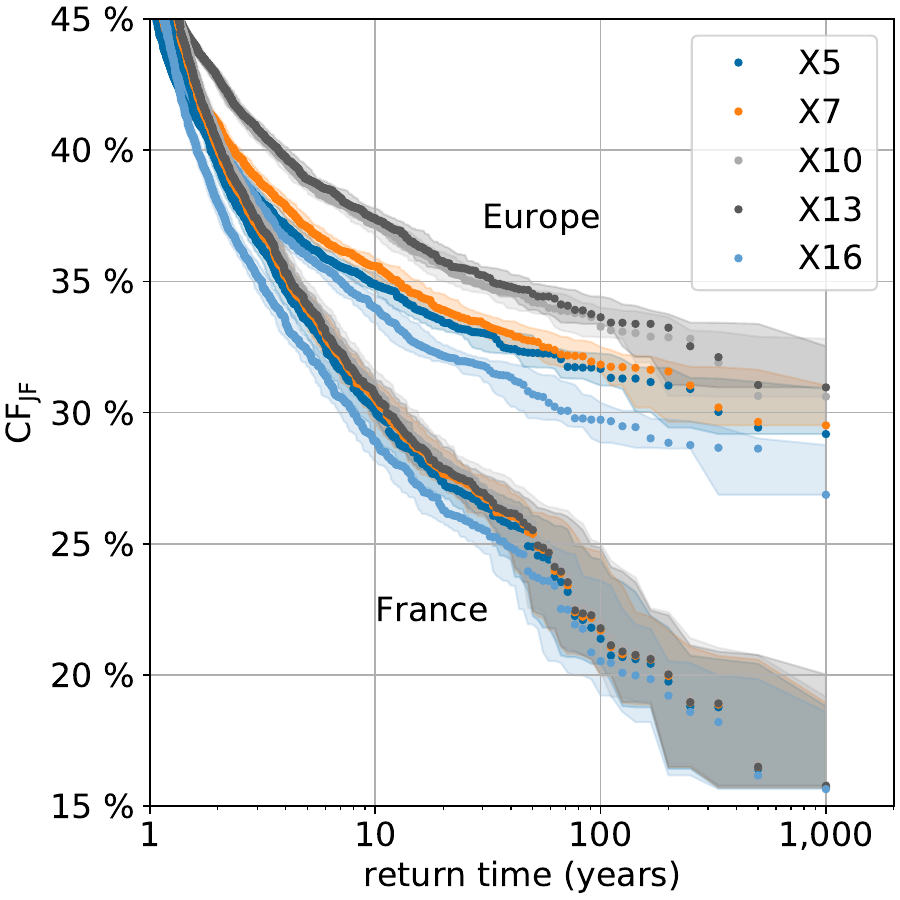}
\includegraphics[width=0.3\textwidth]{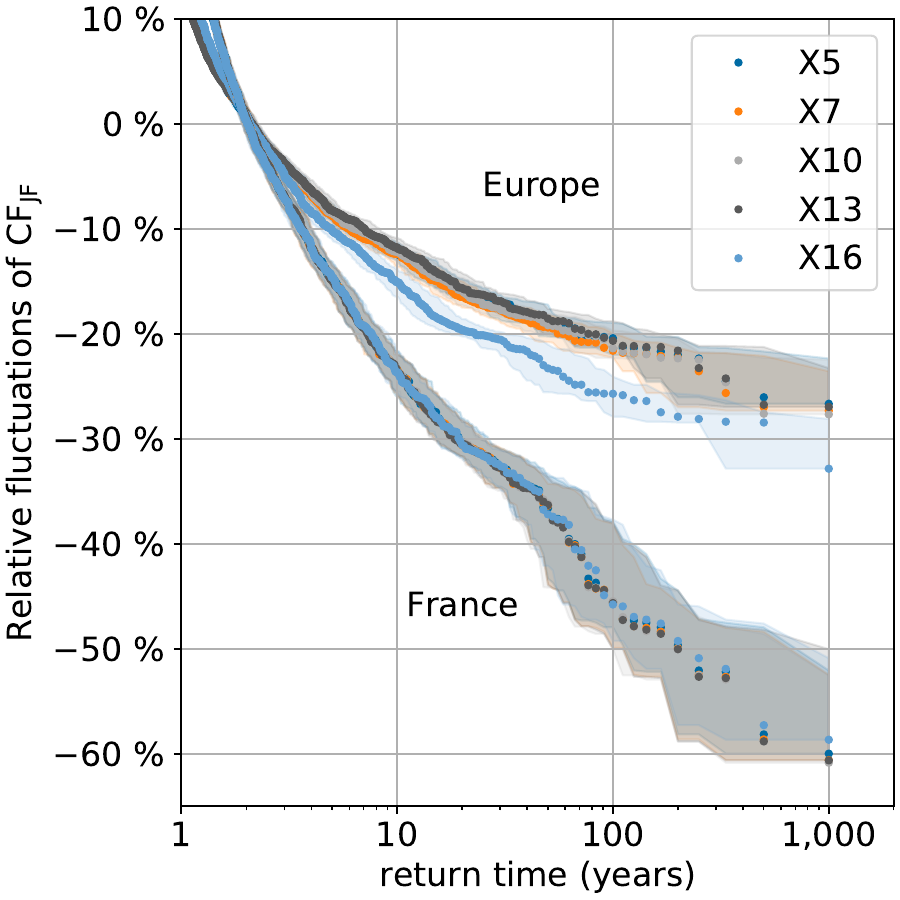}
\includegraphics[width=0.3\textwidth]{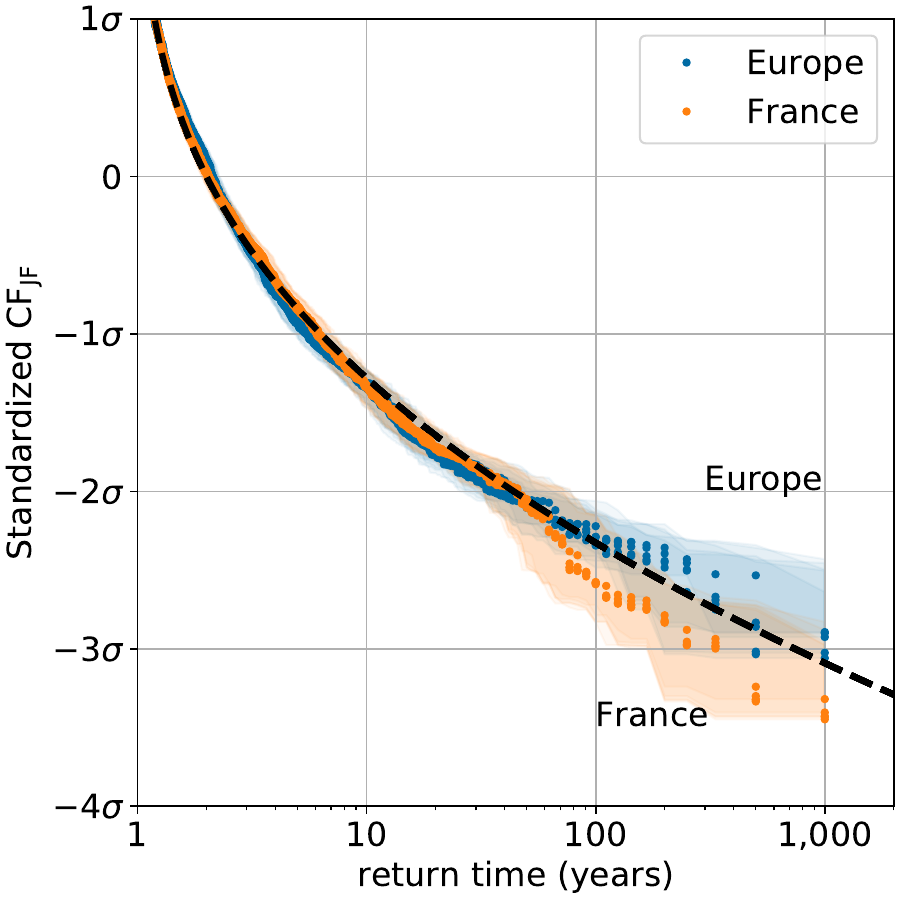}
\caption{Return time curves for low average JF capacity factor $\mathrm{CF}_{\mathrm{JF}}$ (left panel), relative fluctuations of the JF capacity factor $\mathrm{CF}'_{\mathrm{JF}} = (\mathrm{CF}_{\mathrm{JF}} - \langle\mathrm{CF}_{\mathrm{JF}}\rangle)/\langle\mathrm{CF}_{\mathrm{JF}}\rangle $ (middle panel), and standardized JF capacity factor (right panel), see main text for details.
In the first two panels, the 5 scenarios are shown in different colors.
In the last panel, all scenarios for each region are shown in the same color.
The return time curve for a standard normal distribution is shown as a black dashed line.
$95\%$ confidence intervals (shaded) are computed with a bootstrap method (\ref{sec:returntimesappendix}).}
\label{fig:return_time_europe_france_59day}
\end{figure*}

% General properties of the return time curves
First, we consider return time curves for the capacity factor averaged over JF, $\mathrm{CF}_{\mathrm{JF}}$ (figure~\ref{fig:return_time_europe_france_59day}, left panel), for all scenarios, for both France and Europe.
Such curves can be interpreted either as the frequency of occurrence of an event of a given amplitude, or the reverse.
For instance, at the scale of Europe for scenario X16, one may say that a JF capacity factor of $30\%$ has a return time of 100 years (i.e. a 1 in 100 chance of occurring in any given year), or equivalently that the return level at 100 years is $30\%$.
All return time curves have a convex shape.
This means that a simple extrapolation (i.e. a linear fit in figure~\ref{fig:return_time_europe_france_59day}) of return levels for small return times would over-estimate the amplitude of the rarest events.

% Effect of the different scenarios
The return time curves shown in figure~\ref{fig:return_time_europe_france_59day} depend on the geographical area (France or Europe) and scenario considered (less so at the scale of France, where the distribution of installed capacity varies less from one scenario to the other).
For instance, the European JF capacity factor drops below $34\%$ on average once in a decade for scenario X16, but only once in a century for X10 and X13.
However, we show below that these dependencies only come from basic statistical properties (mean and variance of capacity factor) and that events with return times between 10 and 100 years obey universal properties.

First, we show in figure~\ref{fig:return_time_europe_france_59day} (middle panel) return time curves for relative fluctuations of capacity factor $\mathrm{CF}_{\mathrm{JF}}'=(\mathrm{CF}_{\mathrm{JF}}-\langle\mathrm{CF}_{\mathrm{JF}}\rangle)/\langle\mathrm{CF}_{\mathrm{JF}}\rangle$, where $\langle\mathrm{CF}_{\mathrm{JF}}\rangle$ is the climatology.
Unlike absolute capacity factors, return times for fluctuations of capacity factor around the mean depend little on the scenario.
Note, however, that a gap remains for large return times between scenario X16 and the others.
This is due to the larger variance of the capacity factor in this scenario, presumably due to its small amount of offshore wind energy, as shown in the return time curves for standardized capacity factor (i.e. $\mathrm{CF}'_{\mathrm{JF}}/\sigma$ with $\sigma=\sqrt{\left\langle{\mathrm{CF}'_{\mathrm{JF}}}^2\right\rangle}$), where all scenarios collapse (Fig.~\ref{fig:return_time_europe_france_59day}, right panel).
% Effect of spatial scale
In addition, the return time curves of standardized capacity factor for France and Europe also collapse.
Without this normalization (figure~\ref{fig:return_time_europe_france_59day}, left and middle panels), the convexity of the curve is stronger for Europe than for France, which shows that although the return level of typical events (e.g. for 2-year return time, corresponding to the median of the distribution) is similar for both cases, space aggregation decreases significantly the amplitude of rare events.
For instance, on average, the capacity factor in France drops by about $46\%$ once a century, but only by about $20\%$ at the European scale.
This mitigation effect of spatial aggregation on rare events is in fact only due to the reduced variance (\ref{sec:aggregation_statistics}), as figure~\ref{fig:return_time_europe_france_59day} shows (right panel).
The return time curves for standardized capacity factor match well the standard normal distribution (black dashed line), at least below 100-year return times.
We can therefore conclude that the statistics of extreme low JF capacity factor events with return times below 100 years can be assumed to be Gaussian, and all the information about scenarios and spatial scale are encoded in the average and variance.
For rarer events, with return times above 100 years, the uncertainty is larger, and although it seems that events are more extreme than Gaussian at the scale of France and less extreme at the scale of Europe, the available data does not allow to draw definitive conclusions.

%%%%%%%%%%%%%%%%%%%%%%%%%%%%%%%%%%%%%%%%%%%%%%%%

\subsection{Effect of time aggregation on return time curves}

\begin{figure*}[htbp]
\centering
\includegraphics[width=0.49\textwidth]{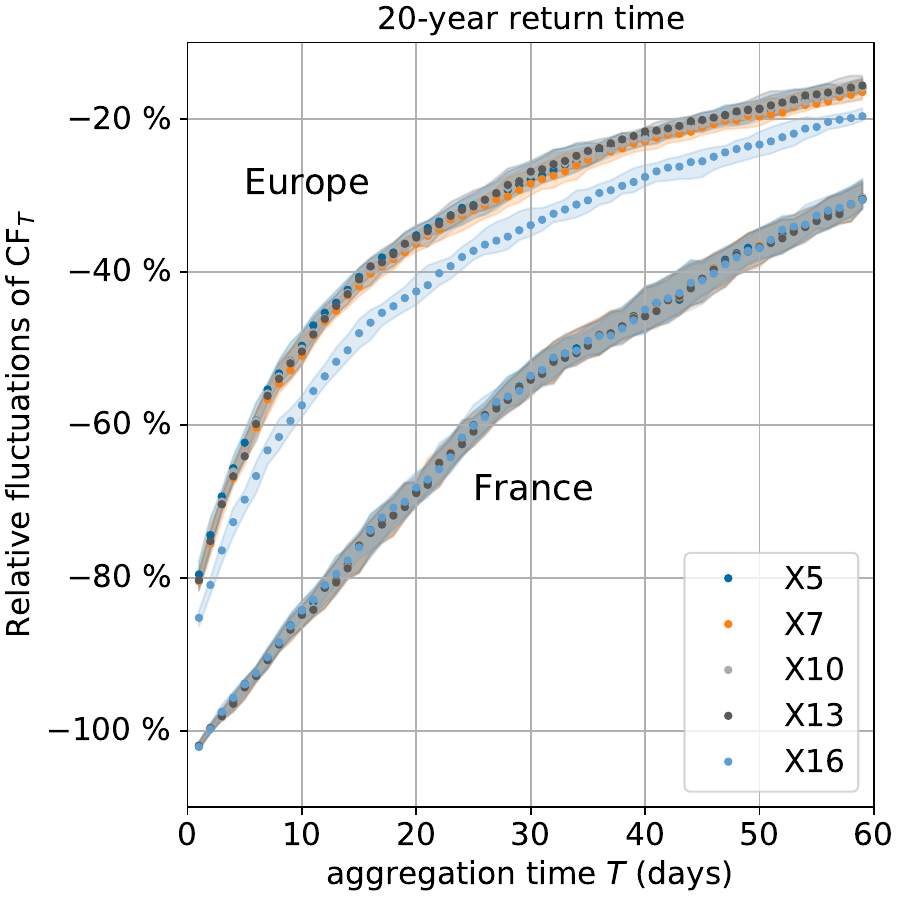}
\includegraphics[width=0.49\textwidth]{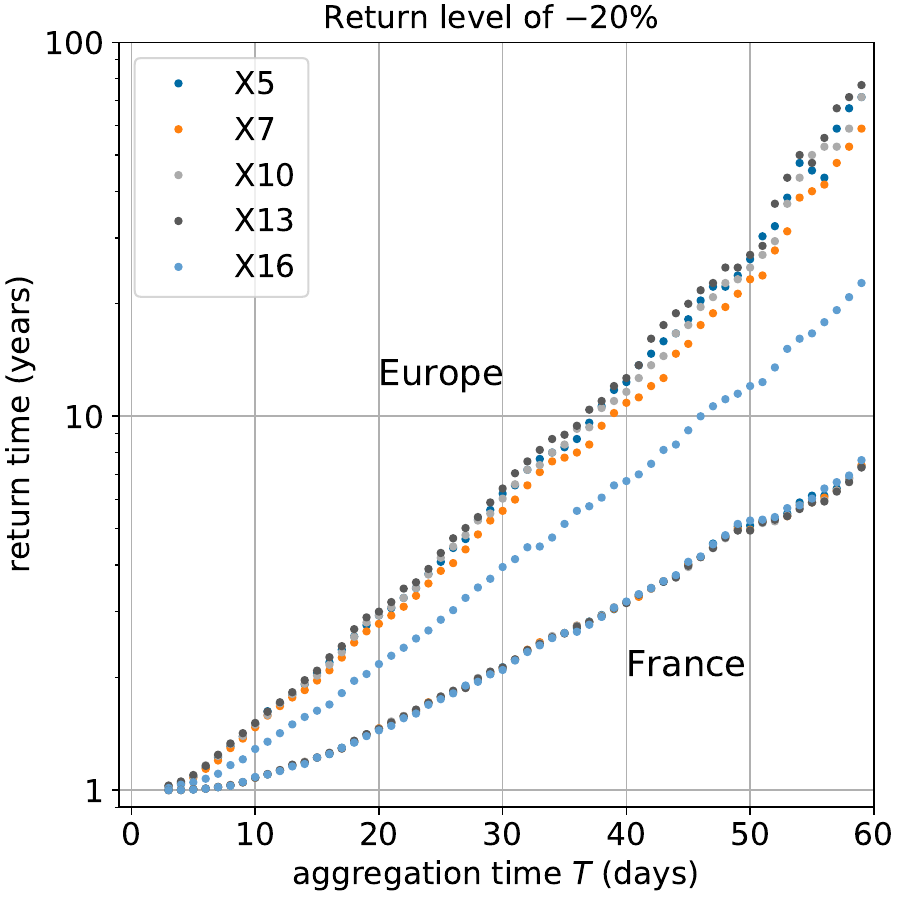}
\caption{Return level $\mathrm{CF}_T'$ of a 1-in-20-year low fluctuation (left panel) and return time of a $\mathrm{CF}_T'=-20\%$ event (right panel), as a function of the aggregation time $T$, for 5 scenarios and for both Europe and France.
For the left panel, the confidence interval is the 95\% confidence interval of the return time curve for a 20 year return time, which is computed as in figure~\ref{fig:return_time_europe_france_59day} for every value of $T$.
}
\label{fig:return_level_vs_T_europe_france}
\end{figure*}

Above we considered extreme events for the JF capacity factor, corresponding to an integration time $T=59$ days (\ref{eq:min_def}).
We now study return times of capacity factor fluctuations for shorter events, down to $T=1$ day.
To do so, we can either fix the return time and plot the return level as a function of the aggregation time $T$ (figure~\ref{fig:return_level_vs_T_europe_france}, left panel; we chose a 20-year return time), or conversely fix the return level and plot the return time as a function of the aggregation time $T$ (figure~\ref{fig:return_level_vs_T_europe_france}, right panel, -$20\%$ return level).

In both panels we note again that the different scenarios lead to very similar statistics, except X16 which yields slightly more extreme events at the scale of Europe, and that spatial aggregation leads to smaller amplitude or less frequent events for Europe compared to France.
We observe a smoothing effect of time aggregation: the amplitude of capacity factor fluctuations, for fixed return time (figure~\ref{fig:return_level_vs_T_europe_france}, left panel) decreases with increasing $T$.
At the European scale, the curve is markedly concave, which means that flexibility measures allowing to smooth energy generation over several days or a few weeks are very efficient (for instance the amplitude of fluctuations is reduced by half when integrating the capacity factor over 20 days), while, by contrast, over longer time scales the amplitude of fluctuations depends less on the aggregation time (integrating from 40 days to 59 days only reduces the fluctuations by a few percents).
A similar behavior has been reported for 10-year wind speed events as a function of their duration~\cite{Leahy2013}.
At the scale of France, these two regimes are less pronounced and the reduction of the amplitude of the fluctuations with the aggregation time is more linear.
This stronger concavity at the European scale means that spatial aggregation is more efficient at reducing the amplitude of rare $\mathrm{CF}_T$ fluctuations with increasing $T$: it is 20\% smaller at the European scale than at the French scale for daily events, while it is about 50\% smaller for 1-month events.

Equivalently, the smoothing effect is seen on the return time curve of capacity factor fluctuations with fixed amplitude (figure~\ref{fig:return_level_vs_T_europe_france}, right panel), which increases approximately exponentially with $T$.

%%%%%%%%%%%%%%%%%%%%%%%%%%%%%%%%%%%%%%%%%%%%%%%%
% Section 4
%%%%%%%%%%%%%%%%%%%%%%%%%%%%%%%%%%%%%%%%%%%%%%%%

\section{Estimating return times with Gaussian processes}\label{sec:gaussian_process}

The analysis carried out in section~\ref{sec:return_times} suggests that extreme capacity factor events behave essentially in a Gaussian manner for long enough aggregation time.
It should be expected, however, that it ceases to be the case for events on shorter time and/or spatial scales.
To investigate the limits of validity of this approximation, we compare the statistics based on capacity factor time series with a simple Gaussian stochastic process, with the same covariance structure.

\subsection{The Gaussian process}

\begin{figure*}%[htbp]
\centering
\includegraphics[width=0.49\textwidth]{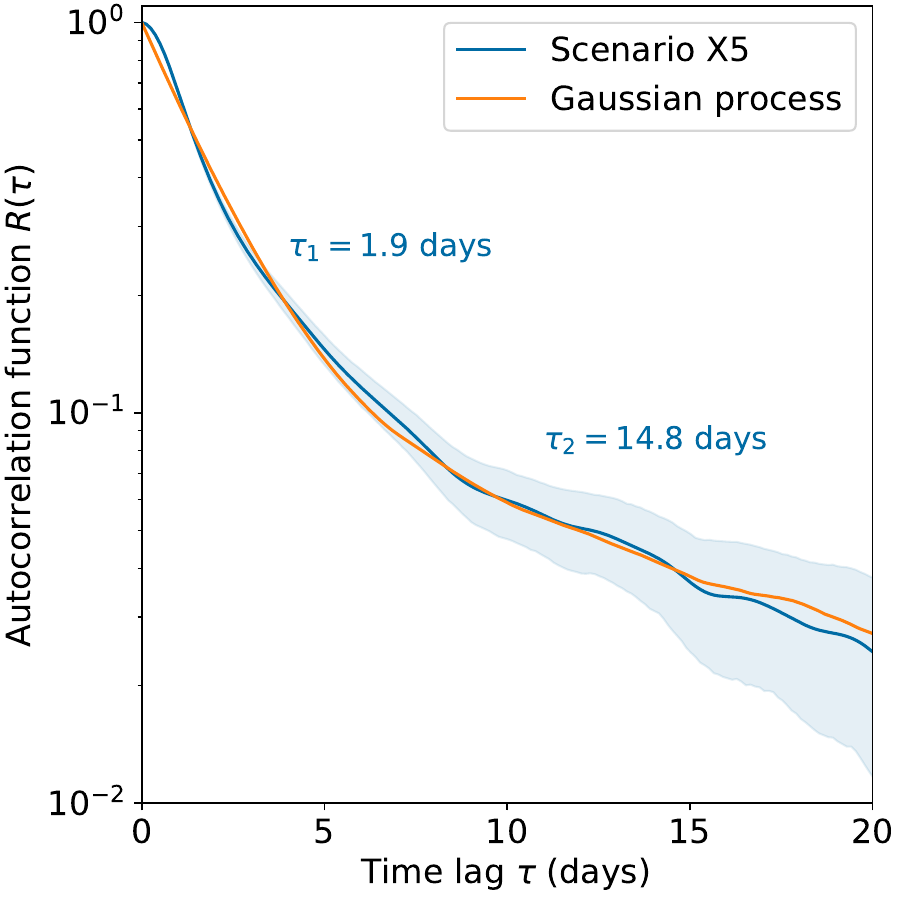}
\includegraphics[width=0.49\textwidth]{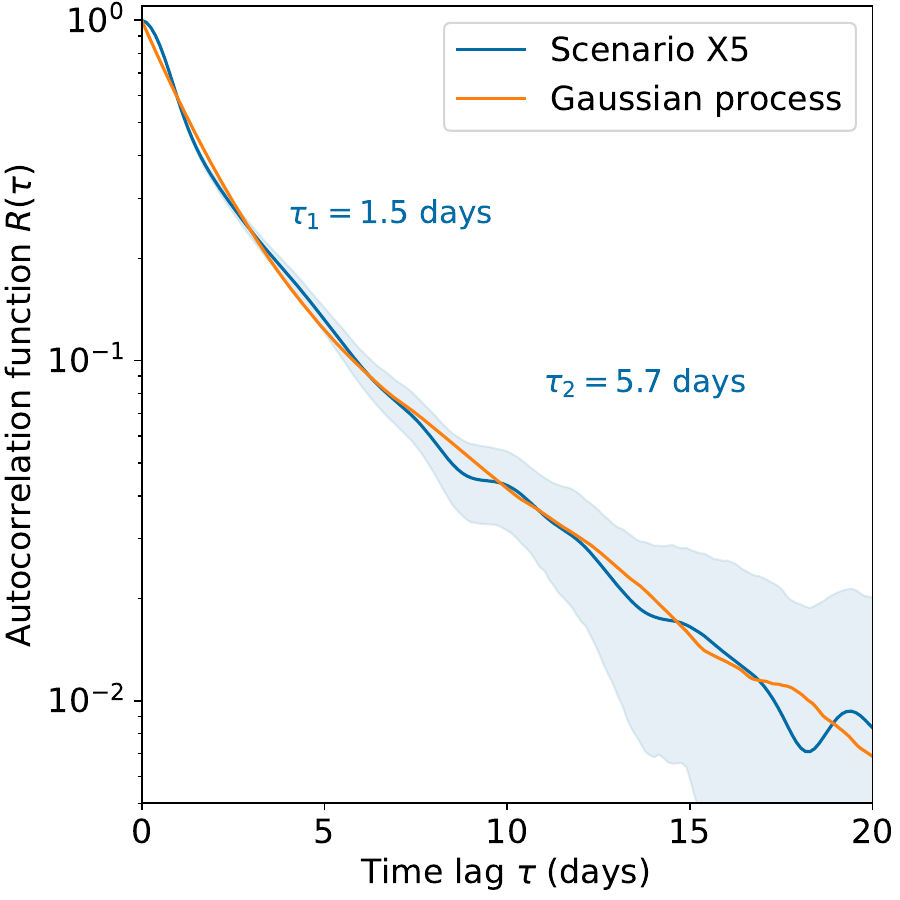}
\caption{Autocorrelation functions of daily capacity factor fluctuations $\mathrm{CF}_{T=1\,\mathrm{day}}'(t)$ for Europe (left) and France (right) for scenario X5 (blue) and the Gaussian process (orange).
Note the logarithmic scale in the y-axis.
The $95\%$ confidence interval (shaded area) is computed with a bootstrap method.
The time scales $\tau_1$ and $\tau_2$ are fitted with a mean square error method.}
\label{fig:autocorelation_function}
\end{figure*}

First we compute the autocorrelation function of the capacity factor fluctuations time series $\mathrm{CF}_{T=1\,\mathrm{day}}'(t)$, daily-averaged to remove the diurnal cycle (figure~\ref{fig:autocorelation_function}).
Because of the seasonality of wind energy production, the autocorrelation function should depend on the calendar time $t_{\mathrm{cal}}$ as well as the time lag $\tau$: $R(t_{\mathrm{cal}}, \tau) = \langle\mathrm{CF}'(t_{\mathrm{cal}})\mathrm{CF}'(t_{\mathrm{cal}}+\tau)\rangle/\langle\mathrm{CF}'(t_{\mathrm{cal}})^2\rangle$.
As the wind energy production statistics vary slowly over the January-February period (section~\ref{sec:cf_seasonal}), we assume that the dependence on the calendar date can be neglected over these two months: $R(t_{\mathrm{cal}}, \tau)\approx R\left(0, \tau\right)$.
We observe that the autocorrelation function (figure~\ref{fig:autocorelation_function}) decays approximately as the sum of two exponentials with characteristic timescales $\tau_1\approx 2$ days and $\tau_2\approx 15$ days for Europe, $\tau_1\approx 1.5$ days and $\tau_2\approx 6$ days for France.
Similar results are obtained for all scenarios.

We therefore construct a simple model for capacity factor fluctuations, as the sum $x_G(t)=x_1(t)+x_2(t)$ of two \emph{Ornstein-Uhlenbeck processes}~\cite[e.g.]{GardinerBook} corresponding to these two time scales:
\begin{equation}
  \begin{array}{ll}
    {\rm d} x_i(t) = - \frac{1}{\tau_i} x_i(t) {\rm d}t + \sqrt{2 D_i} \,{\rm d} W_i(t), & i \in \{1,2\},
  \end{array}
\end{equation}
where $W_1$ and $W_2$ are two independent \emph{Wiener processes}.
This process is Gaussian as the sum of two Gaussian processes.
Its autocorrelation function is a sum of two exponentials which matches well the one obtained from the climate data (figure~\ref{fig:autocorelation_function}).
The noise amplitudes $\{D_i\}_i$ are chosen such that the variance $\tau_1 D_1+\tau_2 D_2$ is equal to the variance of the daily capacity factor, and such that their relative weight $\frac{\tau_2 D_2}{\tau_1 D_1}$ matches the relative weight fitted on the autocorrelation function.
A $59\times10^4$-day long realization of this stochastic process is computed with the \texttt{StochRare} Python module~\cite{Herbert2017}, using a Euler-Maruyama method.

%%%%%%%%%%%%%%%%%%%%%%%%%%%%%%%%%%%%%%%%%%%%%%%%

\subsection{Return time estimates}

\begin{figure*}%[htbp]
\centering
\includegraphics[width=0.5\textwidth]{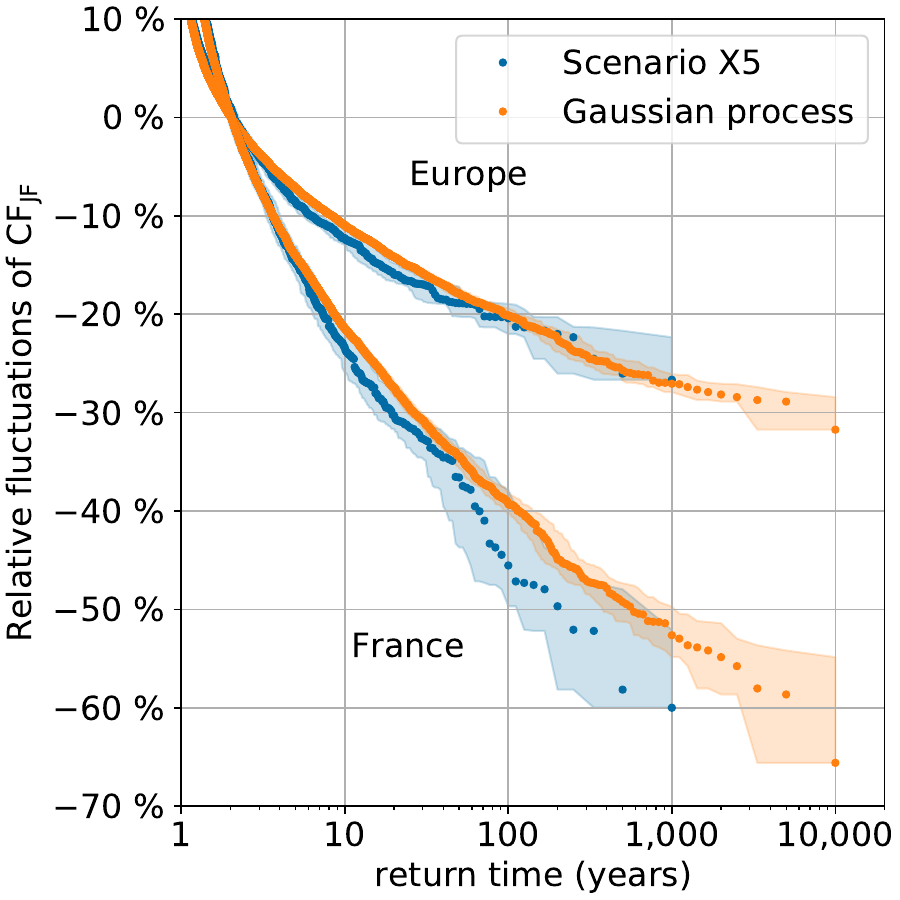}
\caption{Return time curve for low average JF capacity factor fluctuations $\mathrm{CF}_{\mathrm{JF}}'$ for scenario X5 (blue), and for a Gaussian process with the same covariance structure (orange), for Europe and for France. The $95\%$ confidence intervals (shaded areas) are computed with a bootstrap method.}
\label{fig:return_time_OU_europe_X5_59day}
\end{figure*}
Figure~\ref{fig:return_time_OU_europe_X5_59day} compares return time curves for JF capacity factor fluctuations $\mathrm{CF}_{\mathrm{JF}}'$ based on climate data and for the Gaussian process, for Europe and for France.
It shows that the approximation performs well for small return times, below 5 years.
On the other hand, for intermediate return times between 6 and 50 years the Gaussian process lies outside the confidence interval estimated from climate data: the fluctuations are systematically more extreme than the Gaussian process at fixed return time, or more frequent at fixed return level.
For instance, a relative fluctuation of $-16\%$ for Europe ($-30\%$ for France) has a return time of approximately 20 years for climate data and 30 years for the Gaussian process.
Conversely, for Europe, a 1-in-20-year event corresponds to a capacity factor fluctuation of $-16\%$ for climate data and $-14\%$ for the Gaussian process.
For the most extreme events, with return times above 100 years, the Gaussian process is a relatively good approximation at the continental scale but not at the scale of France, where it underestimates the severity or frequency of the events even more than for intermediate return times.
Note however that the uncertainty is larger in this regime.

\begin{figure*}%[htbp]
\centering
\includegraphics[width=\textwidth]{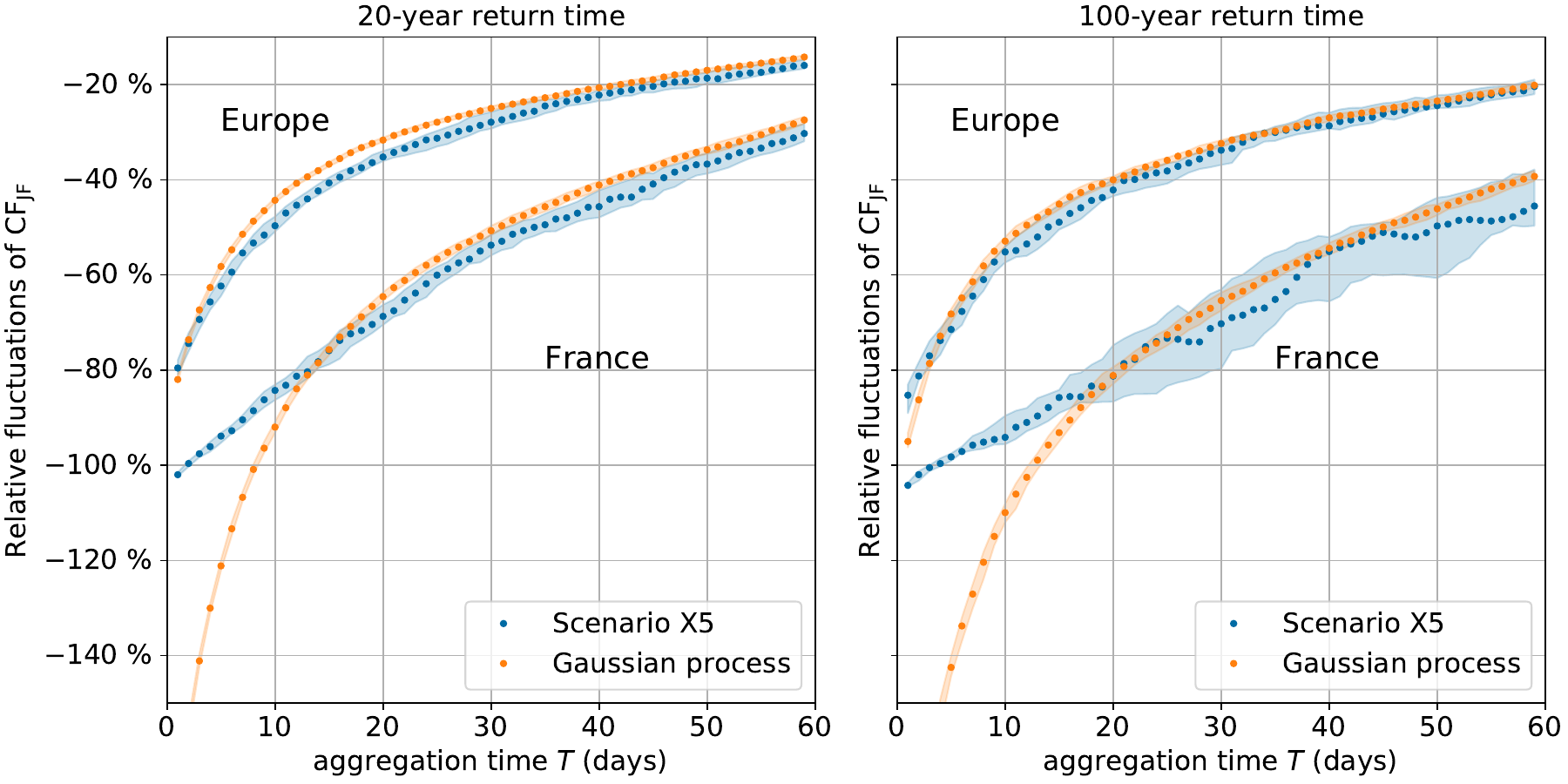}
\caption{Same as figure~\ref{fig:return_time_OU_europe_X5_59day}, but looking at the return level of 1-in-20-year (left panel) and 1-in-100-year events (right panel) for different integration time $T$. The two figures share the same y-axis.}
\label{fig:return_level_vs_T_X5_OU}
\end{figure*}
In figure~\ref{fig:return_level_vs_T_X5_OU}, we compare again the Gaussian process with climate data, but this time, varying the aggregation time, for given return times of 20 years (left panel) and 100 years (right panel).
The above conclusions still hold: overall, the Gaussian process is a better approximation at the scale of Europe than at the scale of France.
In particular, it fails completely for short events (below 15 days) over France.
In the case of Europe, the approximation performs slightly better for rarer events.
In all cases, except 100-year events over France (for which the error bars are much larger), the discrepancy is larger for events lasting between two weeks and a month, and the amplitude of the event is systematically larger than the Gaussian process estimate.
However, relative errors remain relatively small.
For instance, for a two-week event, the 20-year return level is $-42\%$ for climate data and $-38\%$ for the Gaussian process.

%%%%%%%%%%%%%%%%%%%%%%%%%%%%%%%%%%%%%%%%%%%%%%%%

\subsection{Application to reanalysis data}\label{sec:example_reanalysis}

We have shown above that a simple model based on a Gaussian process provides a good approximation of return time curves at the scale of Europe.
One of the main practical interests is that it allows to compute return times for events which are not observed in the original dataset.
Since most datasets are quite short (at most a few decades), this is a very interesting feature to study rare events.
Here we apply this method to the Renewables.Ninja dataset~\cite{Staffell2016}, which provides hourly wind capacity factors for European countries, %EU-27 plus United-Kingdom, Norway and Switzerland.
computed with the MERRA-2 reanalysis product over the period 1980-2016.
It addresses some of the limitations of our climate data, such as an increased spatial resolution and the use of bias-correction with respect to available data from national Transmission System Operators.
We use it here as a proof of concept for the method.
\begin{figure*}%[htbp]
\centering
\includegraphics[width=0.5\textwidth]{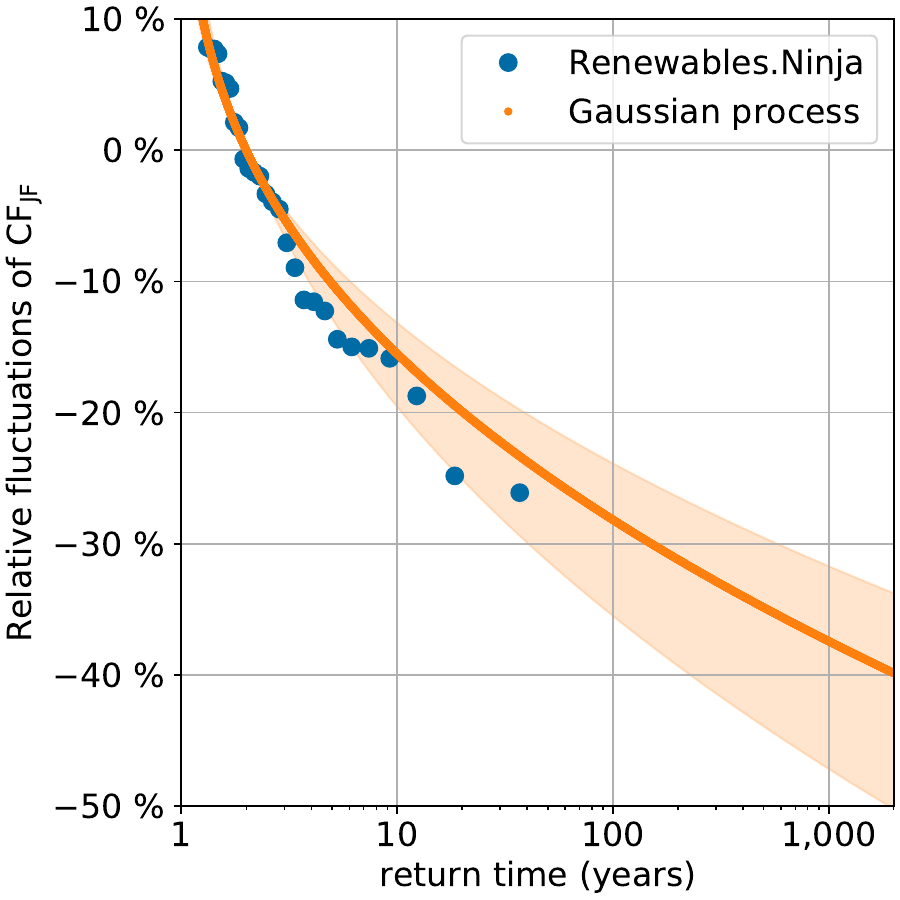}
\caption{Return time curve for low JF capacity factor fluctuations $\mathrm{CF}_{\mathrm{JF}}'$ for the Renewables.Ninja dataset estimated directly (blue) and using a Gaussian process with the same covariance structure (orange).
The 95\% confidence interval is computed from the uncertainty in the estimate of the correlation time $\tau_2$, due to the short length of the data (see~\ref{sec:estimate_tau_2}).
}
\label{fig:return_time_Ninja}
\end{figure*}
Figure~\ref{fig:return_time_Ninja} shows the return time curve for JF capacity factor fluctuations over Europe estimated directly from the dataset, and its extension to arbitrarily rare events using a Gaussian process.
Because the dataset is short, there is some uncertainty in the estimate of the second correlation time $\tau_2$, which propagates to the return time estimates with the Gaussian process (\ref{sec:estimate_tau_2}).
Again, the Gaussian process estimate matches well the direct return time curve estimate when data is available.
While the installed capacity in this dataset (as of 2017 for version v1.1) is different from the scenarios considered above, the return time curves are very similar, in agreement with our finding in section~\ref{sec:return_times}.
This approach allows us to extrapolate return times for rare events: for instance, for a 100-year event, we expect relative fluctuations of $-28\%$.
For this scenario where the total installed capacity is $\mathrm{IC}_{\mathcal{D}}=110$~GW and the average JF capacity factor is $\langle\mathrm{CF}_{\mathrm{JF}}\rangle = 31\%$, a 100-year return time event represents an average loss of 9.5~GW over 2 months, or a loss of 13.5~TWh.

%%%%%%%%%%%%%%%%%%%%%%%%%%%%%%%%%%%%%%%%%%%%%%%%
% Conclusion
%%%%%%%%%%%%%%%%%%%%%%%%%%%%%%%%%%%%%%%%%%%%%%%%

\section{Conclusion}

The question we have addressed here is the estimation of the probability and amplitude of events where wind energy production remains low for periods ranging from a few days to the entire January-February period.
Using a very long climate simulation, we have computed return time curves for low capacity factor events over Europe or France using different scenarios of installed capacity up to 1000-year return times.
% Methodological aspect. Gaussian approximation
A first result is that the return times of extreme winter capacity factor depend on the geographic area and installed capacity scenario only through the average and standard deviation of the capacity factor; after standardization they follow a universal curve.
This simplifies the discussion of projections of extreme events of wind energy production, as the properties of the various scenarios can be obtained by simple rescaling.
We further show that the return time curves can be estimated using a simple Gaussian process with the same covariance as the capacity factor time series generated from the climate model.
This approximation performs well at the scale of Europe, although events tend to be slightly more extreme than Gaussian, especially at intermediate return times (on the order of 10 years) and durations (from a week to a month).
The same remark applies in a magnified way at smaller spatial scale (France), where the quality of the approximation degrades significantly for events with return times above 100 years or durations below two weeks.
Nevertheless, for long events at the scale of Europe, this provides a method to estimate return times even when observational or model data is scarce.
We illustrate this by constructing a return time curve based on a capacity factor dataset constructed from more precise data, using reanalyses, but much shorter than our climate data.
This return time curve can be extended to arbitrarily rare events.

% Effect of space and time aggregation. Short vs long events.
Finally, we have discussed the effect of space and time aggregation for rare events.
Our results show that spatial aggregation significantly reduces the amplitude of extreme events: for daily events, the capacity factor fluctuations in Europe is about 20\% lower than in France, and for 1-month events it is about 50\% lower, at 20-year return times.
Similarly, we have shown that longer events are much less extreme than short ones: less than 20\% loss compared to the average capacity factor for two-month events on a European scale, 40\% for two weeks and 80\% for one day.
At the scale of France, these figures become 30\% for two-month events, 70\% for two weeks and 100\% for one day.
Nevertheless, in absolute terms the wind energy shortfall for two-month events is on the order of 20-59~GW for Europe and 3-15~GW for France, depending on the scenario installed capacity, for 20-year return times, and about 20\% more for 100-year return times.
Our results, obtained with a simple energy model, therefore suggest that such long events might put important constraints on future energy systems, given the fact that flexibility means become scarcer on such long timescales.
This provides an incentive for studying rare, long-lasting events in more comprehensive energy models.

\ack
This work was granted access to the HPC/AI resources of CINES under the allocation 2018-A0050110575, 2019-A0070110575 and 2020-A0090110575 made by GENCI.
Data analysis was carried out using the resources of the \emph{Centre Blaise Pascal} at ENS de Lyon.
We are grateful to Emmanuel Quemener for his help with the platform.
This research has been partially funded by the \emph{Institut des Mathématiques pour la Planète Terre (IMPT)}.
We would like to thank Laurent Dubus for his very valuable comments on a preliminary version of this manuscript.

%%%%%%%%%%%%%%%%%%%%%%%%%%%%%%%%%%%%%%%%%%%%%%%%
% Appendix
%%%%%%%%%%%%%%%%%%%%%%%%%%%%%%%%%%%%%%%%%%%%%%%%

\appendix

\section{Additional information on the wind energy model}\label{sec:windmodelappendix}

The $0.9^{\circ}\times 1.2^{\circ}$ grid of CESM is remapped with a bilinear interpolation method to the $0.5^{\circ}\times 0.5^{\circ}$ grid\footnote{Available on the Copernicus Climate Data Store \href{https://cds.climate.copernicus.eu}{https://cds.climate.copernicus.eu} [accessed in October 2020]}  used by the ECEM project~\cite{Troccoli2018}.
This allows a better resolution of e-Highway2050 `` cluster '' regions (section \ref{sec:scenarios}).
Similarly to studies using reanalysis data~\cite{Cannon2015, Ohlendorf2020}, no bias correction is applied to the wind speed, in order to preserve the self-consistency of the climate model output.
Furthermore, such bias correction methods have been validated for average wind speeds but not for extreme values, and we can expect some degree of error smoothing due to the large scales considered here.

To compute the wind speed at turbine height $\mathrm{WS}$, we choose a hub height of $h_{\mathrm{hub}}= 100\,\mathrm{m}$ for both onshore and offshore turbines, and assume a power law profile with exponent $\alpha=\frac{1}{7}\approx 0.143$~\cite{Karman1921,Jerez2015}:
$\mathrm{WS} = \mathrm{WS}_{\mathrm{10 m}} \left(\frac{h_{\mathrm{hub}}}{h_{\mathrm{ref}}}\right)^{\alpha}$, where $\mathrm{WS}_{\mathrm{10 m}}$ is the wind speed at the reference height $h_{\mathrm{ref}}= 10\,\mathrm{m}$.

\begin{figure}[htbp]
  \centering
  \includegraphics[width=\linewidth]{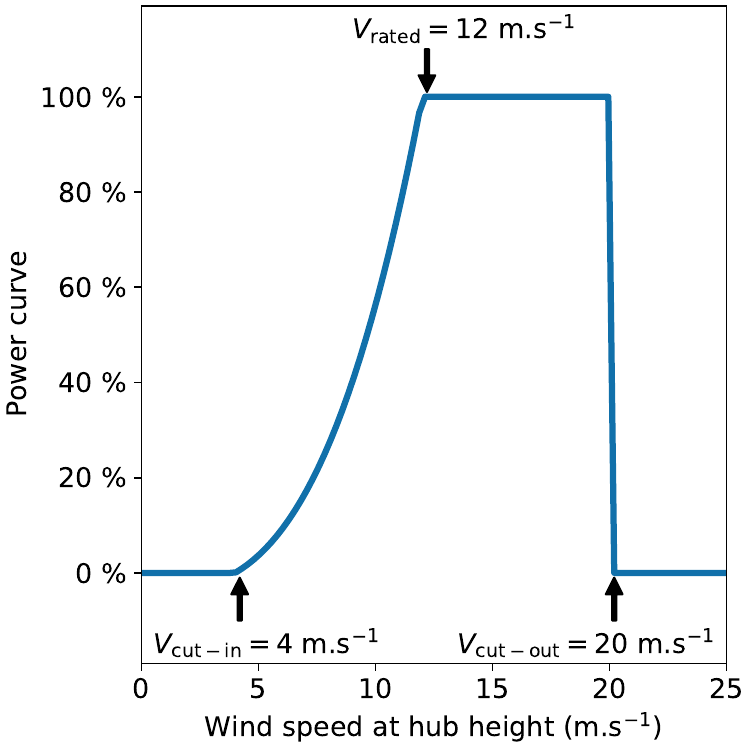}
  \caption{Power curve used in the energy model.}
  \label{fig:powercurve}
\end{figure}
To relate the generated power to the wind speed, we use a simple power curve (figure~\ref{fig:powercurve}) with a cubic increase of the generated power after some cut-in value, until the rated power is reached.
The turbine stops after a cut-off wind speed for security reasons.
The equation for the power curve is:
\begin{equation}
\mathrm{PC} = \left\{
    \begin{array}{llllll}
    0 & \mathrm{if} & & & \mathrm{WS} & < \mathrm{V}_{\mathrm{ci}}  \\
    \frac{\mathrm{WS}^3 - \mathrm{V}_{\mathrm{ci}}^3}{\mathrm{V}_{\mathrm{r}}^3 - \mathrm{V}_{\mathrm{ci}}^3} & \mathrm{if} & \mathrm{V}_{\mathrm{ci}} & \leq & \mathrm{WS} & < \mathrm{V}_{\mathrm{r}}  \\
    1 & \mathrm{if} & \mathrm{V}_{\mathrm{r}} & \leq & \mathrm{WS} & < \mathrm{V}_{\mathrm{co}}   \\
    0 & \mathrm{if} & \mathrm{V}_{\mathrm{co}} & \leq & \mathrm{WS} & \\
    \end{array}
\right.
\label{eq:power_curve}
\end{equation}
The parameters ($\mathrm{V}_{\mathrm{ci}}=4$ m.s$^{-1}$, $\mathrm{V}_{\mathrm{r}}=12$ m.s$^{-1}$, $\mathrm{V}_{\mathrm{co}}=20$ m.s$^{-1}$) are chosen to correspond to a VESTAS V110-2.0MW turbine\footnote{From \url{https://thewindpower.net/turbine\_en\_590\_vestas\_v110-2000.php} [accessed in October 2020]} as done in the European Climatic Energy Mixes (ECEM) project~\cite{Troccoli2018}.

\begin{figure}[htbp]
  \centering
  \includegraphics[width=\linewidth]{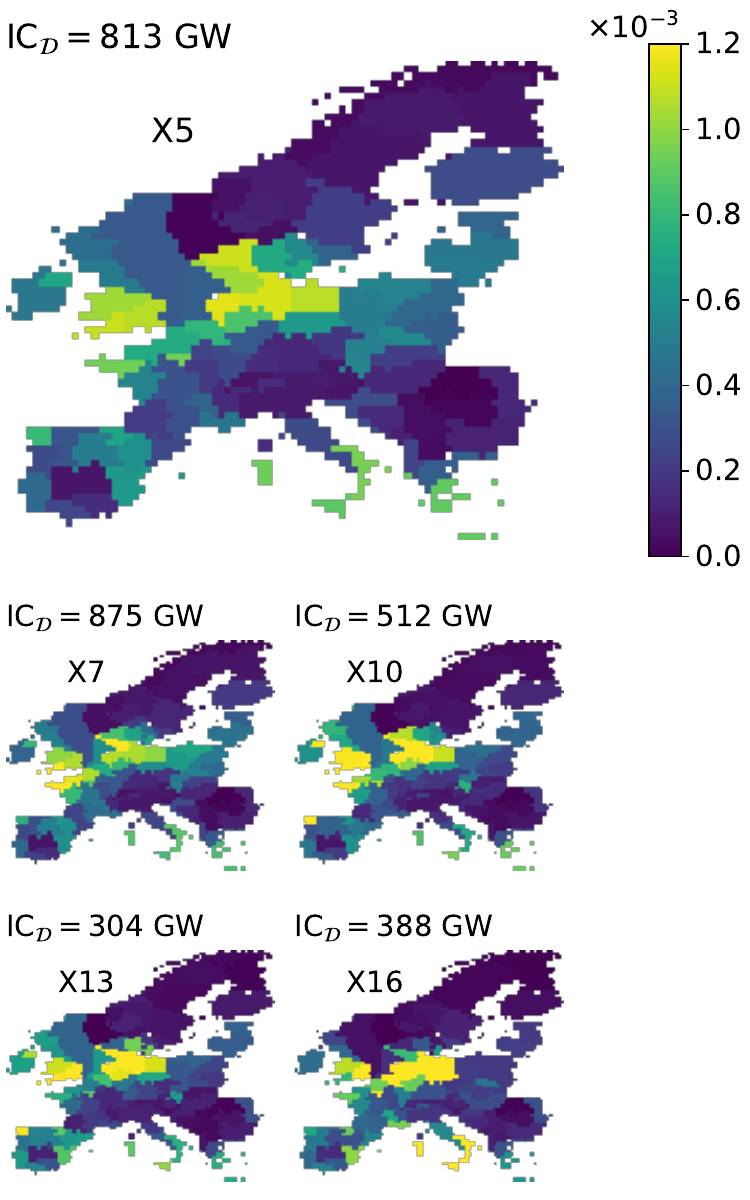}
  \caption{Spatial distribution of installed capacity $\frac{\mathrm{d}x\mathrm{d}y\,\mathrm{IC}(x,y)}{\mathrm{IC}_{\mathcal{D}}}$ (in $\%$) for all 5 e-Highway2050 scenarios. The total installed capacity $\mathrm{IC}_{\mathcal{D}}$ is shown at the top left of each panel.}
  \label{fig:spatial_distribution_IC_eHighway}
\end{figure}

The total capacity factor $\mathrm{CF}(t)$ over a region $\mathcal{D}$ in equation~(\ref{eq:CF}) depends on the scenario only through its spatial distribution $\frac{\mathrm{d}x\mathrm{d}y\,\mathrm{IC}(x,y)}{\mathrm{IC}_{\mathcal{D}}}$.
Installed capacities in the e-Highway2050 project are defined at the scale of about a hundred ``clusters'' built from European NUTS-3 regions (Fig.~\ref{fig:spatial_distribution_IC_eHighway}).
This allows for a greater spatial granularity than scenarios defined at the scale of countries such as ENTSO-E's Ten-Year Network Development Plan scenarios~\cite{TYNDP2022} or European Commission's reference scenarios~\cite{REF2020}.
Figure~\ref{fig:spatial_distribution_IC_eHighway} shows the spatial distribution of installed capacity at this cluster scale for all 5 e-Highway2050 scenarios.
Contrarily to onshore clusters, e-Highway2050 does not provide a geographical description of equivalent ``cluster'' regions in the North Sea.
Therefore, we use a dataset constructed for offshore energy~\cite{Saint-Drenan2020} to manually assign maritime regions to offshore wind capacities.

In the main text we focus on properties which are common to all scenarios, but the distribution of installed capacity has a clear effect both on the annual (as reported in section~\ref{sec:cf_annual}, it varies from 25.5\% to 28.9\%) and seasonal mean capacity factor (Fig.~\ref{fig:mean_std_annual_CF}).
Indeed, the mean winter capacity factor varies from one scenario to the other between $39\%$ and $41.7\%$.
This difference is much smaller than the winter-average of the daily standard deviation (from $10.9\%$ for X5 to $13.6\%$ for X16, see figure~\ref{fig:mean_std_annual_CF}), but of the same order of magnitude as the inter-annual variability, computed as the standard deviation of the mean winter capacity factor (from $2.9\%$ for X5 to $3.6\%$ for X16, not shown).
Small differences in the mean winter capacity factor leads to significant differences in term of total energy production when they are integrated over the whole winter: for instance a $37\%$ mean capacity factor corresponds to a total winter energy production $7.5\%$ lower than a $40\%$ mean capacity factor.

Some scenarios have specific characteristics: X5 and X7 have a higher capacity factor in summer than X16, but a similar capacity factor in winter, while X10 and X13 always have a capacity factor larger than all other scenarios.
In scenario X16, where the share of offshore is by far the lowest, the daily standard deviation is at least 2 percentage points higher than in all other scenarios, and the inter-annual standard deviation is at least 0.6 percentage points higher than in all other scenarios.

%%%%%%%%%%%%%%%%%%%%%%%%%%%%%%%%%%%%%%%%%%%%%%%%

\section{Illustration of the effect of aggregation on capacity factor statistics}\label{sec:aggregation_statistics}

In this section, computations are made for scenario X5, but the conclusions also hold for the other scenarios.

\begin{figure*}[htbp]
\centering
\includegraphics[width=\textwidth]{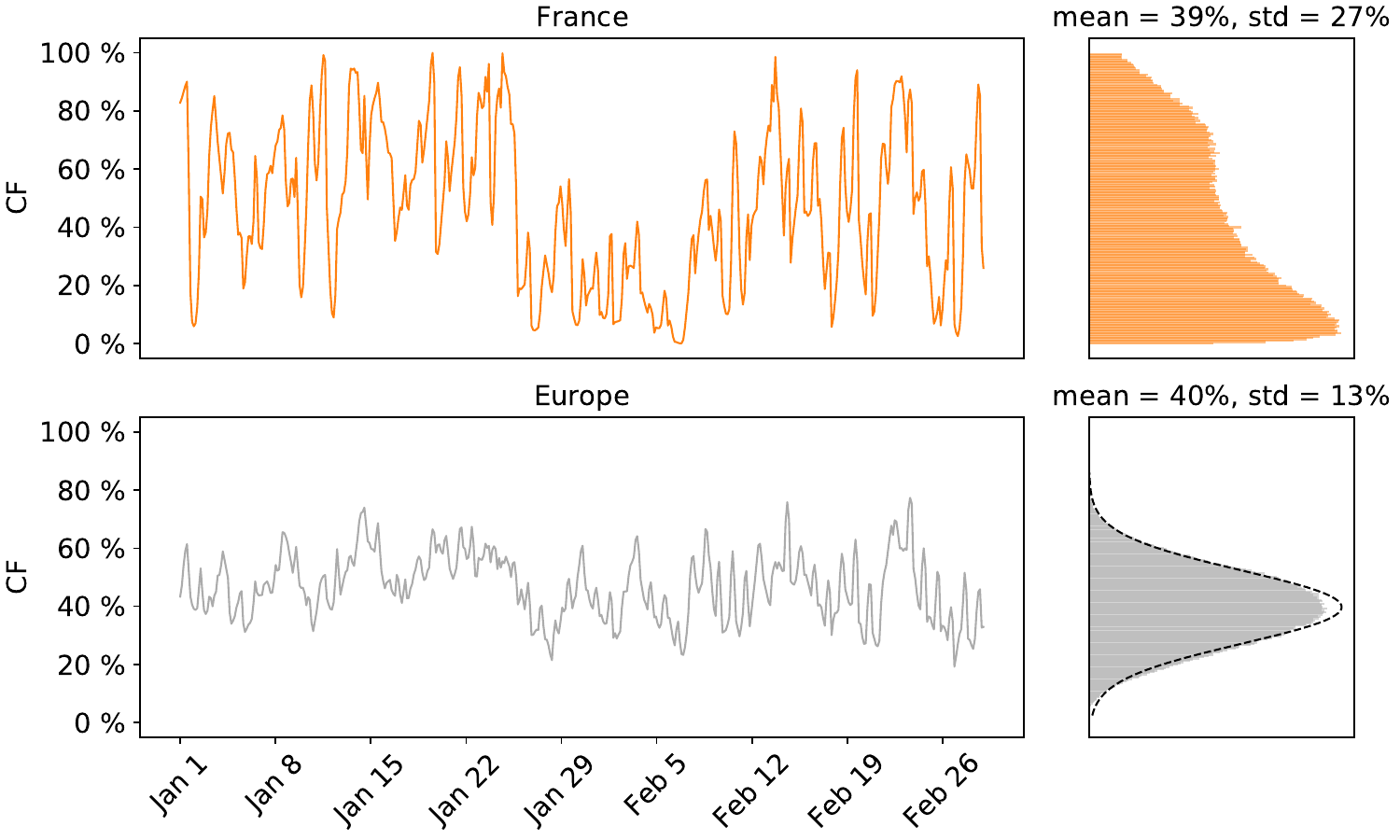}
\caption{Time series of the 3-hourly winter capacity factor for year 989 of the climate simulation (left), and probability density function (PDF) of the 3-hourly winter capacity factor computed on the 1000 years of data (right), at the scale of France (top row) and Europe (bottom row), for scenario X5.
The mean and the standard deviation of the winter capacity factor are indicated above each PDF.
A Gaussian fit of the PDF is shown in black dashed line.}
\label{fig:space_aggregation_illustration}
\end{figure*}

On figure~\ref{fig:space_aggregation_illustration}, we compare two time series from our simulations of the 3-hourly capacity factor in winter and its probability density function (PDF) aggregated at the national scale (taking the example of France) and at the continental scale (Europe).
At the scale of France, the capacity factor undergoes strong fluctuations, with daily ramp-up and ramp-down events of about $80\%$.
The PDF of the capacity factor is strongly skewed: low-production events are much more frequent than high-production events.
A long lasting low production event is observed from late-January to early-February in the sample time series shown in figure~\ref{fig:space_aggregation_illustration}.
Hence, at this spatial scale the smoothing effect of space aggregation appears to be limited by the correlations of the wind field.
On the other hand, at the European scale, the PDF of winter capacity factor is nearly Gaussian.
While the average capacity factor is almost the same in both cases, the standard deviation ($13\%$) is about half the one at the scale of France ($27\%$), illustrating the strong smoothing effect of space aggregation at the continental scale.
The persistent low production event in France is no longer visible at the European scale.

\begin{figure*}[htbp]
\centering
\includegraphics[width=\textwidth]{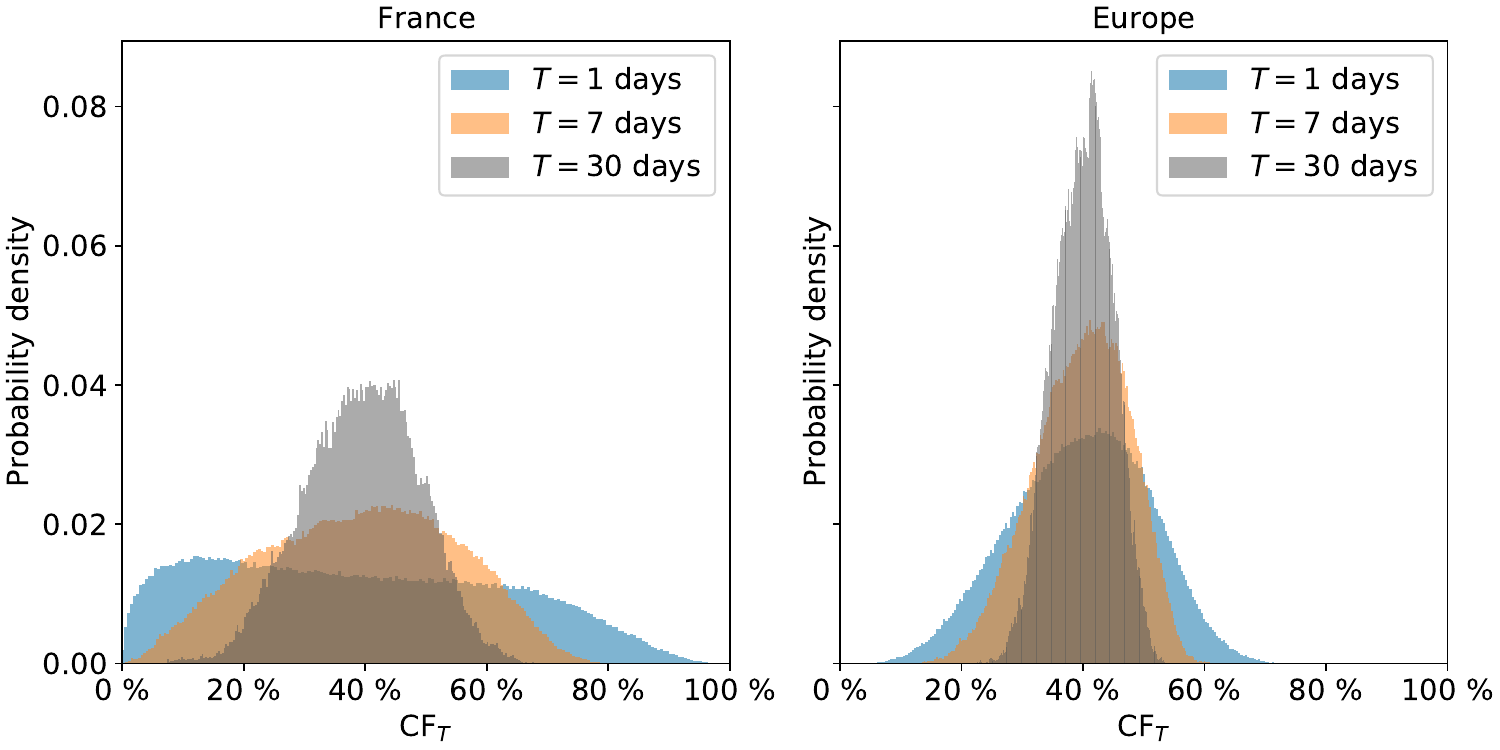}
\caption{Probability density function of the moving average of the capacity factor $\mathrm{CF}_T(t)$ in winter, for $T=1$ day, $T=7$ days and $T=30$ days, at the scale of France (left) and Europe (right), in scenario X5.
}
\label{fig:pdf_CF_europe_france_T}
\end{figure*}

To characterize the impact of the timescale on the statistics of the fluctuations, we show in figure~\ref{fig:pdf_CF_europe_france_T} the PDF of the moving average of the capacity factor $\mathrm{CF}_T$ (\ref{eq:CF_T}) over France and Europe, for three integration times: $T=1$, 7 and 30 days.
At the scale of France, the main effect of time aggregation is that it makes the PDF of the time-averaged capacity factor more and more Gaussian as the length of the averaging window increases.
This effect is less spectacular at the European scale, as the capacity factor statistics are already closer to Gaussian at small integration times, due to the effect of spatial aggregation.
In both cases, the variance decreases significantly: it reduces by a factor of $2.5$ for France and $2.3$ for Europe when the integration time goes from $T=1$ day to $T=30$ days.

\section{Return time estimates}\label{sec:returntimesappendix}

We compute return times for seasonal minima of capacity factor aggregated over a time interval $T$.
More precisely, for a stochastic process $\{X(t)\}$ in a time interval $[T_1,T_2]$, we define the random variable
\begin{equation}
  a_T = \min_{t\in[T_1, T_2-T]} \left\{ \frac{1}{T} \int_t^{t+T} \mathrm{d}u\,X(u) \right\},
  \label{eq:min_def}
\end{equation}
where $T$ is the integration (or aggregation) time.

Here, $X$ is a periodic stochastic process, with period $T_p=1$ year, $T_1$ is the first time step of January and $T_2$ is the last time step of February.
We assume that the correlation time of the process is much smaller than the 1-year periodicity of $\{X(t)\}$, such that the samples of the random variable $a_T$ obtained from all the winters in the climate simulation can be considered statistically independent.
We thus have $N=1000$ independent samples, i.e. one event per climate year.
This independence assumption is reasonable with the climate model used here, but the approximation would degrade if long-term memory effects, for instance due to the ocean, were included in the model.

In section~\ref{sec:return_times}, the stochastic process $\{X(t)\}$ alternatively stands for the capacity factor $\mathrm{CF}_T$ itself, its relative fluctuations $\mathrm{CF}'_T = \left(\mathrm{CF}_T - \langle\mathrm{CF}_T\rangle \right) / \langle\mathrm{CF}_T\rangle$, and the standardized fluctuations $\mathrm{CF}'_T/\sigma$ with $\sigma=\sqrt{\left\langle{\mathrm{CF}'_T}^2\right\rangle}$ the standard deviation.

We define the return time $r(a)$ of the return level $a$ as the average time to wait before seeing a year with $a_T \leq a$.
It can be easily estimated as follow.
Let us define the random variable $Y(a)$ by $Y(a)=1$ if $a_T<a$ and $Y(a)=0$ otherwise.
By construction, $Y(a)$ has a binomial distribution with parameter $F(a)=\mathbb{E}[Y(a)]$, the cumulative distribution function of $a$:
\begin{equation}
  F(a) = \mathbb{P}[a_T \leq a] = \int_{-\infty}^a \mathbb{P}[a_T=a'] \,\mathrm{d}a',
\end{equation}
which can be estimated with $\hat{F}(a) = \hat{n}(a)/N$, where $\hat{n}(a)$ is the number of samples of $a_T$ lower than the threshold $a$.

We note $\tau(a)$ the random variable which is the number of years to wait before seeing an event with $a_T<a$.
The probability of $\tau(a)=nT_p$ is related to the negative binomial distribution for one failure.
Indeed, the probability $\mathbb{P}[\tau(a)=T_p]$ that the event $a_T<a$ occurs within one year is $F(a)=\mathbb{P}[a_T \leq a]$, and the probability to see this event at year $n$ is $\mathbb{P}[\tau(a)=nT_p]=\left(1-F(a)\right)^{n-1}F(a)$.
The return time is thus $r(a)=\mathbb{E}[\tau(a)] = \frac{T_p}{F(a)}$.
A direct estimator is therefore
\begin{equation}
  \hat{r}(a) = \frac{N T_p}{\hat{n}(a)}.
\end{equation}
This choice of extreme events gives a natural definition of the return time.
For example, an event $a_{2\mathrm{y}}$ with a return time of two years ($r(a_{2\mathrm{y}})=2 T_p$) is defined such that $F(a_{2\mathrm{y}})=\frac{1}{2}$, which means $a_{2\mathrm{y}}$ is the median of the distribution: one half of the events are above $a_{2\mathrm{y}}$ and the other half is below $a_{2\mathrm{y}}$.
Return times are defined in a similar way by~\cite{Leahy2013}, who computed them for wind speed at several sites in Ireland using short observational records.

In practice, to compute a return time curve, we first sort the $N$ samples in increasing order $\{a^k_T\}_{k\in\{1,\cdots,N\}}$ such that the number of times $a_T$ is lower than $a^k_T$ is its rank in the sorted sample: $\hat{n}(a^k_T)=k$.
Then, we simply plot $\left\{a^k_T\right\}_{k\in\left\{1,\cdots,N\right\}}$ on the $y$-axis and $\left\{\frac{NT_p}{k}\right\}_{k\in\{1,\cdots,N\}}$ on the $x$-axis, as done in~\cite{Lestang2018}.

We calculate the $95\%$ confidence interval by applying a non-parametric bootstrap method~\cite{Wilks2011,DelSole2022}.

%%%%%%%%%%%%%%%%%%%%%%%%%%%%%%%%%%%%%%%%%%%%%%%%

\section{Estimating return times with short datasets}\label{sec:estimate_tau_2}

Using a single time scale $\tau_1$ gives a poor approximation of the autocorrelation function and thus a poor estimate of the return time curve (not shown).
Therefore it is necessary to correctly estimate the second time scale $\tau_2$ if one wants to make prediction on the return time curve of extreme low JF capacity factor fluctuations $\mathrm{CF}_{\mathrm{JF}}'$.

There are two types of errors associated with a limited amount of data: the error on the return time curve directly computed with the data available (the `` direct error ''), and the error on the fit of $\tau_2$ which propagates when making an estimation of the return time curve with a Gaussian process (the `` $\tau_2$-error '')

\begin{figure*}%[htbp]
    \centering
    \includegraphics[width=0.49\textwidth]{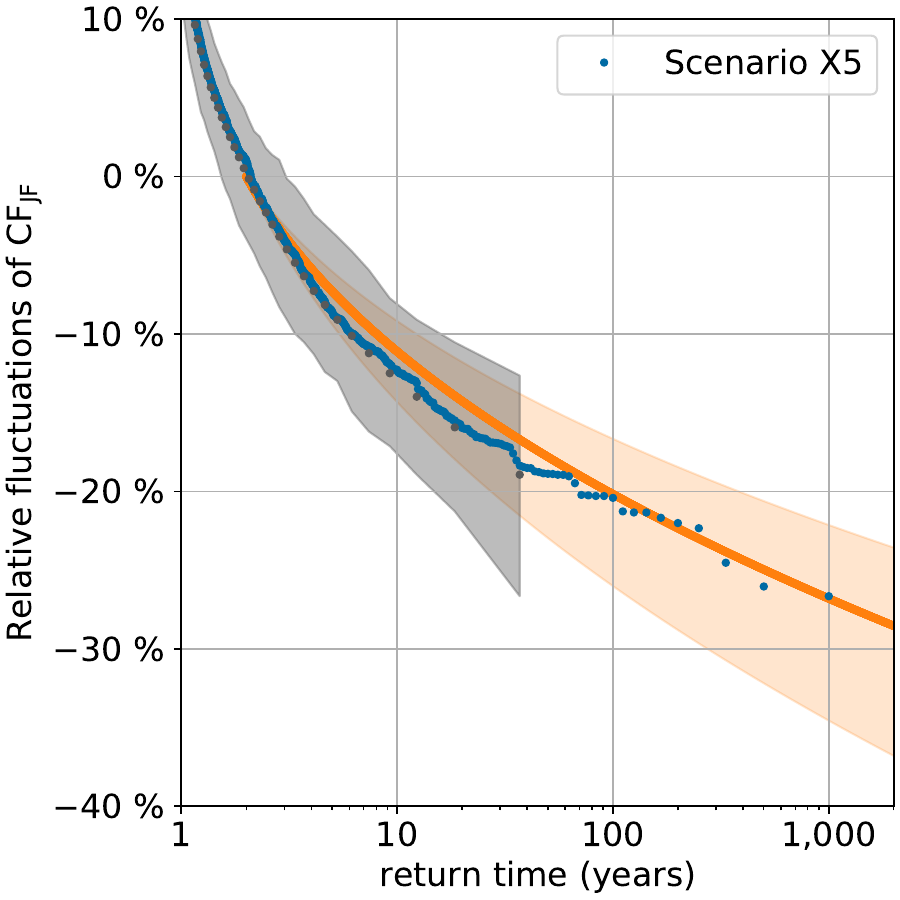}
    \caption{Errors on the return time curve estimate due to lack of data: (grey) error due to the direct effect of data size (`` direct error '', see the main text) and (orange) error on the sampling of $\tau_2$ (`` $\tau_2$-error '', see the main text). For both errors, 95\% confidence intervals are shown. The same return time curve for the JF capacity factor fluctuations $\mathrm{CF}_{\mathrm{JF}}'$ as in figure~\ref{fig:return_time_OU_europe_X5_59day} is shown in blue.}
    \label{fig:return_time_2_confidence_intervals}
\end{figure*}

Both errors are computed with a bootstrap method by drawing randomly a large number (typically $10^3-10^4$) of 37-year-long bootstrap samples.
Each bootstrap sample is used to estimate directly a return time curve (for the direct error), or to estimate $\tau_2$ and compute the return time curve of the associated Gaussian process (for the $\tau_2$-error).
In figure~\ref{fig:return_time_2_confidence_intervals}, the 95\% confidence interval is shown in grey for the direct error and orange for the $\tau_2$-error.
The method based on a Gaussian process yields smaller errors than a direct estimate.

In section~\ref{sec:example_reanalysis}, we use the $\tau_2$-error found in our climate simulation to estimate the 95\% confidence interval for the 37-years-long Renewables.Ninja dataset.

\newpage
\bibliographystyle{iopart-num}
\bibliography{biblio}

\end{document}